\documentclass{mem}
\usepackage{natbib}
\usepackage{graphicx}
\usepackage[a4paper]{hyperref}
\idline{00 }{00 }
\begin{document}
\title{
Evidence for GR rotational frame-dragging in the light from the Sgr A* supermassive black hole
}

   \subtitle{}

\author{
B. \,Aschenbach\inst{1} 
          }

  \offprints{B. Aschenbach}

\institute{
Max-Planck-Institut f\"ur extraterrestrische Physik, Postfach 1312, 85741 Garching, Germany
}

\authorrunning{Aschenbach }

\titlerunning{Mass and spin of the Sgr A* black hole}

\abstract{
The analysis of flare start-times confirms the periods found years ago 
(Aschenbach et al., 2004) in the near-infrared and X-ray light-curves 
related to the Sgr A* black hole. 
The assignment of the frequencies found to radial and vertical epicyclic frequencies 
$\nu\sb{\rm r}$, $\nu\sb{\rm v}$, respectively, 
as well as to 
the Kepler orbital frequency $\nu\sb{\rm K}$ reveals resonances of $\nu\sb{\rm v}$/$\nu\sb{\rm r}$=7:2, and 
$\nu\sb{\rm K}$/$\nu\sb{\rm v}$=3:1. 
The highest observed frequency of 10~mHz  is identified as the Kepler frequency corrected 
by the rotational frame-dragging frequency, as expected from the  Lense-Thirring effect. 
These frequency assignments conclude a black hole mass 
of M~=~(4.10--4.34)$\times$10$\sp{6} M_\odot$
 and a spin of a~=~(0.99473--0.99561). 
}
\maketitle{}

\section{Introduction}

The center of our Galaxy is suspected to contain a supermassive black hole.
Observationally this conjecture is based on the analysis of
motions and orbit sizes of stars moving around some central mass (Sch\"odel et al. 2002; Ghez et al. 2005; 
Gillessen et al. 2009).
These observations suggest the presence of a fairly small volume containing a mass of several million solar masses. 
Mass and spin of the black hole have also been
estimated from quasi-periodic oscillations 
(Abramowicz et al. 2004; Aschenbach et al. 2004; Aschenbach 2004) 
 which have been suggested to be present
in the near-infrared 
(Genzel et al. 2003)
and X-ray (Aschenbach et al. 2004)
 light-curves of Sgr A*.

These candidate periods are suspect because of their enhanced power spectral density in at least three Sgr A* 
observations, which 
include 
the October 26, 2000 ({\it{Chandra}}, Baganoff et al. 2001), the October 3, 2002 ({\it{XMM-Newton}},
Porquet et al. 2004) and the near-infrared (NIR) observations of June 16, 2003 
(Genzel et al. 2003).
These observations
were particularly interesting because they showed fairly large outbursts of Sgr A*. I have summarized the results, 
and
I have proposed candidate
period-like structures, centered around 110~s, 219~s and 1173~s (Aschenbach et al. 2004).
Later measurements, both in the NIR and the X-ray band, were not conclusive on these periods.
In most of the cases there was no indication of a period at all, but when a time structured signal was suggested by the data, 
it happened to be close to
the candidate period of 1100~s. This includes a  1330~s period 
(B\'elanger et al. 2006),
 and a 
 $\sim$1500~s period (Meyer et al. 2008).
However, each one of 
these observations, taken as a single event, was not and cannot be considered  
to be of that
significance which would justify a statistically unambiguous claim of the detection of a period.
But the fact that these coincidences among the light-curve observations exist as far as
the putative periods are concerned, is encouraging further study.

The indication of more than one period in the light-curve data suggests the possibility that
the light-curve is not dominated by just one period but that the signal is actually modulated 
by one ore even more frequencies.
So far, the analyses had to deal with light-curves which are usually subject to a large amount of noise (be it white, or red 
or even pink).
Therefore I took the opportunity to analyse the starting times of a sequence of X-ray flares which were observed 
with {\it{XMM-Newton}}
between 31 March and 5 April 2007, which in principle provide an independent access to periodic patterns.

\section{Data and analysis}

In a recent paper Porquet et al. (2008)
 reported for the first time
a high-level X-ray flaring activity of Sgr A* observed with
{\it{XMM-Newton}} between 31 March and 5 April 2007.
Five flares in a row were detected. Their start-times were assessed 
in a quantitatively justified  way by Porquet et al. (2008). They  
were
determined to  291913503, 292051530, 292073530, 202084630, 292092330 in seconds relative to the
on-board clock readings of {\it{XMM-Newton}}.
The shortest separation between any two of the flares is 7700~s,
and 178827~s is the widest interval.
Given these times I checked each interval between any two of the flares
 for accommodating, as closely as possible, a natural number of trains of a trial period covering
the range from 10~s to 8000~s with a spacing of  1~ms, i.e. $\sim$8$\times$10$\sp 6$ periods were checked.
The data provide 5 time-settings, or four independent time-intervals. The choice of four independent  
time-intervals is, however,  somewhat arbitrary; one can use the time difference between two consecutive  
events or one can choose
the time difference measured against a reference, i.e. the first flare, the second flare, etc. I decided to take 
all possibilities
into account, which for five flares makes 10 measurements. It is obvious that these 10 time-interval measurements 
are not independent of each other,
but this procedure tends to reduce any biases in the measurements, e.g. large measurement errors for the one or 
other start-time. The problem with this
approach is that this procedure is likely to produce not only some solutions but also  combinations of them, because 
of the
oversampling of the information. But any eventual results can be screened for such events and eliminated afterwards.

The algorithm (see {\it{Appendix}} 1) looks for a regular flare start-time pattern, keeping in mind
 that most of the flares
have possibly a brightness below the detection limit and are missed in the observations. The quantitative 
search determines the minimum
 residual between the observed interval and the time of either n-times or (n+1)-times of the trial period.
The values of the residuals are squared and added for the 10 available
intervals, divided by the number of measurements (10) and after taking the square root they end up as 
root-mean-square (rms) value, which is then divided by the trial period. In other words, the mean relative rms-mismatch 
per measurement  between observation and trial period is calculated.  
This is a dimensionless quantity, and this quantity can take values between 0 and 0.5.
The inverse of that quantity I call 'goodness of fit' (GOF), which is the ratio of the trial period and the 
rms deviation from this trial
period for the 10 time-intervals. 
This is NOT a significance test, it is just a search algorithm for suspects!
The 'goodness of fit' versus the trial period, which I call a periodogram, is shown in
Fig. 1. 
\begin{figure*}[t!]
\resizebox{\hsize}{!}{\includegraphics[angle=270,scale=.80]{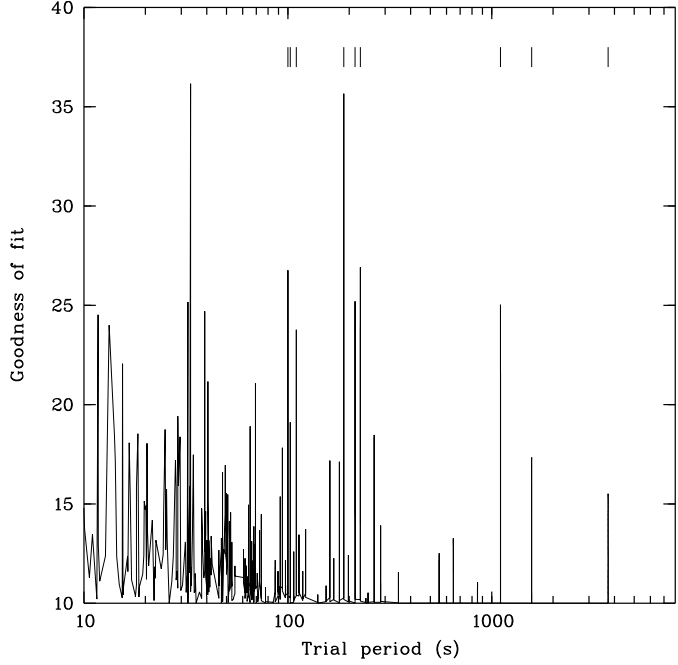}}
\caption{\footnotesize
Periodogram of the trial periods computed from the X-ray flare start-times of the
April 2007 {\it{XMM-Newton}} observation. The ticmarks in the uppermost
part of the graph delineate the positions or periods discussed in detail in the text.
Data with a 'goodness of fit' value of less than 10 are not shown. For the
definition of 'goodness of fit' see the text.
}
\end{figure*}
\begin{table*}[t!]
\caption{Periods in the X-ray flare start-times
of the April 2007 {\it{XMM-Newton}} observations.
\label{tbl-1}}
\begin{center}
\begin{tabular}{cccccc}
\hline
\\
position     &  period P  & line width $\Delta$P/P  & frequency $\nu$  &  mode assign.    \\
             &  (s)       & (FWHM)                  &    (mHz)         &  \\
\hline
\\
      1      & 100.016    & 1.2$\times$10$\sp{-4}$  & 9.9984           & $2\nu\sb{\rm{6}}+\nu\sb{\rm{7}}+\nu\sb{\rm{9}}$\\
      2      & 102.777    & 1.7$\times$10$\sp{-4}$  & 9.7298           & $2\nu\sb{\rm{6}}+\nu\sb{\rm{7}}$\\
      3      & 109.980    & 1.5$\times$10$\sp{-4}$  & 9.0926           & $2\nu\sb{\rm{6}}+\nu\sb{\rm{9}}$\\
      4      & 188.046    & 1.7$\times$10$\sp{-4}$  & 5.3178           & $\nu\sb{\rm{6}}+\nu\sb{\rm{7}}$\\
      5      & 213.650    & 2.6$\times$10$\sp{-4}$  & 4.6806           & $\nu\sb{\rm{6}}+\nu\sb{\rm{9}}$\\
      6      & 226.656    & 2.6$\times$10$\sp{-4}$  & 4.4120           & \underline{$\nu\sb{\rm{6}}$}\\
      7      & 1103.893   & 1.4$\times$10$\sp{-3}$  & 0.90589          & \underline{$\nu\sb{\rm{7}}$}\\
      8      & 1569.126   & 2.8$\times$10$\sp{-3}$  & 0.63730          & $\nu\sb{\rm{7}}-\nu\sb{\rm{9}}$\\
      9      & 3723.179   & 7.5$\times$10$\sp{-3}$  & 0.26859          & \underline{$\nu\sb{\rm{9}}$} \\
\\
\hline
\end{tabular}
\end{center}
\end{table*}

The periodogram shown in Fig. 1 reveals nine prominent peaks which are listed in Table 1. 
The third column of Fig. 1 shows the width of the outstanding periods, which is very small, indicating that a search with 
a 1~ms stepping is too wide for any periods shorter than $\sim$30~s. For the range with trial periods shorter than 35~s 
the screening was repeated with a step size of 10$\mu$s. The corresponding periodogram is shown in Fig.~2.

\begin{figure*}[t!]
\resizebox{\hsize}{!}{\includegraphics[angle=270,scale=.80]{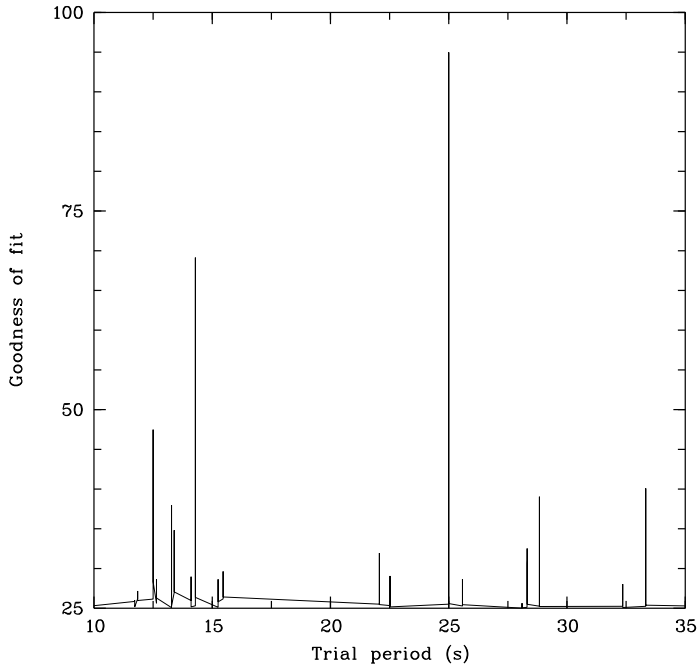}}
\caption{\footnotesize
Periodogram of the trial periods computed from the X-ray flare start-times of the
April 2007 {\it{XMM-Newton}} observation in the range
of 10~s to 35~s using a screening of 10~$\mu$s steps.
Data with a 'goodness of fit' value of less than 25 are not shown.
}
\end{figure*}

The peak at 33.3321 re-appears, and other peaks at shorter periods show-up. 
These large peaks, however, have in common that the periods are exactly (to the fourth 
digit) given by P$\sb 1$/n with n a natural number. They do not provide independent 
information, but none of the periods with P$>$100~s, listed in Table 1, can be 
represented as a simple linear relation between any one of the listed periods. 

However, when analysed in the frequency domain linear relations become apparent for P$>$100~s, and there are just three independent 
frequencies left, out of which the rest can be made up by linear combinations. 
Because the search algorithm is in the period domain rather than in the frequency domain these three frequencies provide 
independent information.
 
The nine periods consist of two groups with one group centered around $\sim$106~s and a second group centered around 
$\sim$214~s, and  two periods at $\sim$1104~s ($\nu\sb 7$) and $\sim$3723~s ($\nu\sb 9$), respectively, as well as their difference 
frequency ($\nu\sb 8$). 
Both, the group of the three frequencies around $\sim$106~s and the group of the three frequencies around $\sim$213~s each 
follow exactly the frequency relations for side band frequencies created by  $\nu\sb 7$ and 
$\nu\sb 9$.  In principle, there are eight side band frequencies for each 
group and its central frequency. Apparently, only three of these nine possible frequencies show up, for each group, with a high peak in the periodogram. 
The central frequencies $\nu\sb{\rm c}$ of the two frequency groups are calculated to 
$\nu\sb{\rm{c,1}}$=9.3611~mHz (P$\sb{\rm{c,1}}$=106.825~s) and $\nu\sb{\rm{c,2}}$=4.68055~mHz (P$\sb{\rm{c,2}}$=213.65~s). 
Obviously, $\nu\sb{\rm{c,2}}$=$\nu\sb{\rm{c,1}}$/2, i.e., there is just one fundamental frequency, which is 
$\nu\sb{\rm{c}}$=$\nu\sb{\rm{c,1}}$.   
 In summary, the observationally fundamental four flare time-intervals have provided 
three independent frequencies, i.e. $\nu\sb{\rm{c}}$ (and $\nu\sb{\rm{c}}$/2), 
$\nu\sb{\rm 7}$ and $\nu\sb{\rm 9}$.

Of course, I have run a number of simulations assuming that five flares happen to occur at random over the entire 
observation time
of 230~ks. GOF values as high as observed are found, for 
 one or another period, quite frequently for periods less than about 100~s, but they 
tend to occur much less frequent
with increasing period, so that the periods close to 1000~s and above are outstanding.

In addition, the periods showing up in the periodogram derived from the flare start-times are very much 
reminiscent of the periods suggested earlier to be present in the Sgr A* light-curves because of their 
enhanced power spectral density, i.e. the observations of October 26, 2000 ({\it{Chandra}}), 
October 3, 2002 ({\it{XMM-Newton}}) and the NIR flare on June 16, 2003
(for a summary see
Aschenbach et al., 2004). The numbers to be compared are 106~s (110~s), 214~s (219~s) and 1104~s
(1173~s). The numbers without brackets are derived from the analysis of the flare start-times taken in 2007, 
and the number in brackets 
are from the light-curves taken before 2004.

In view of this, 
I have not undertaken the tremendous task of calculating the probability that the observed periods are not random 
events, because
that probability is definitely not zero. Any associated  significance number just helps to decide whether to continue 
or stop the investigations.
Because of the astonishing overlap between the periods with those, which show some enhanced power spectral density in the flares mentioned 
above,  it is useful to investigate the possibility that the periodic structures are associated
with the Sgr A* black hole.

\section{Analysis and results}

The periodogram presented  suggests the existence of three independent frequencies.  
Lacking any alternative model, which invokes more than just one frequency, I restrict the following analysis
to an attempt of assigning the observed frequencies to matter oscillations somewhere in an accretion disk which is bound 
by a rotating black hole. 
I suggest to consider that a hot spot is created from a pre-existing fluctuation by resonant waves traveling 
through the accretion disk. These waves are in the radial direction, transporting mass and energy from the 
outskirts to the black hole with a frequency $\nu\sb{\rm r}$, and in the vertical direction, mainly 
raising mass density and 
energy density, with a frequency $\nu\sb{\rm v}$. The amplification of a fluctuation is most effective if the 
waves get into resonance, i.e. $\nu\sb{\rm v}$/$\nu\sb{\rm r}$=n:m, with n and m natural numbers, and m as low as possible, 
preferentially m=1, or m=2. 
The resonance effect becomes even more dramatic if one or both of the waves get into resonance with the 
orbital motion of the initial fluctuation, which is the Kepler frequency $\nu\sb{\rm K}$, i.e. 
$\nu\sb{\rm K}$/$\nu\sb{\rm v}$=k:l, with k and l natural numbers, and l as small as possible, preferentially 
l=1 or l=2. Such a resonance can also be considered to hold for 
$\nu\sb{\rm K}$ and $\nu\sb{\rm r}$. 
The presence of parametric resonances between the epicyclic frequencies has originally been suggested by
Klu\'zniak and Abramowicz (2001a,b) to explain the 3:2 high frequency quasi-periodic oscillation pairs, which are
observed in a number of low mass X-ray binaries most likely containing a black hole companion.

In the following I use the letters P for a period, $\nu$=1/P for the associated frequency and $\Omega$ for the  
 frequency expressed in GR units, i.e. 
c=G=M=1 (c is the speed of light, G is the gravitational constant and M is the mass of the black hole). 
$\nu$ and $\Omega$ are related through $\Omega$/$\nu$=2$\pi$GM/c$\sp 3$. 

In their 1973 paper Cunningham and Bardeen studied "The optical appearance of a star orbiting an extreme 
Kerr black hole. Recognizing that ".. the 'photon' trajectory can loop around the black hole any number of times', 
they concluded that there will be more than one image of the 
 orbiting star seen by the infinitely remote observer, which are separated 
in time by one 'photon' orbit, and they called these images 'direct image', 'one-orbit image', 'two-orbit' image, etc. 
The effect on the observer is that the modulation of the light does not follow the Kepler orbital frequency but it is modified 
by the circular 'photon' orbit frequency $\Omega\sb{\rm{ph}}$ such that 
$\Omega\sb{\rm l}$/$\Omega\sb{\rm K}$=(1-$\Omega\sb{\rm K}$/$\Omega\sb{\rm{ph}}$)$\sp{-1}$. $\Omega\sb{\rm l}$ is the frequency 
with which the light appears to be modulated in the observer's frame. 
$\Omega\sb{\rm{ph}}$ is the same frequency as the rotational frame-dragging frequency (see {\it{Appendix}} 2). 
As long as $\Omega\sb{\rm K}$/$\Omega\sb{\rm{ph}}$$\ll$1,  $\Omega\sb{\rm l}\approx \Omega\sb{\rm K}$, but if this condition 
is violated and the frequency ratio gets closer to one, there is a large difference and $\Omega\sb{\rm{l}}$ can be 
boosted up to several times of $\Omega\sb{\rm{K}}$. This is the case when the black hole has a spin of ${\it a} \approx$ 1, and if 
the Kepler orbit radius 
is just a few gravitational radii or less. If this happens, the modulation of the light-curve 
is no longer with $\Omega\sb{\rm{K}}$, and there 
is no frequency at the value of $\Omega\sb{\rm{K}}$. The by far highest frequency is then $\Omega\sb{\rm{l}}$.

The frequency data shown in Table 1 suggest that $\nu\sb{\rm r}$=$\nu\sb{\rm 9}$, 
$\nu\sb{\rm v}$=$\nu\sb{\rm 7}$ and $\nu\sb{\rm l}$=$\nu\sb{\rm c}$. Using the relativistic expressions 
for $\Omega\sb{\rm{ph}}$, $\Omega\sb{\rm{K}}$, $\Omega\sb{\rm{v}}$ and $\Omega\sb{\rm{r}}$ (see {\it{Appendix}} 2)  
 a best-fit reveals P$\sb{\rm K}$=1/$\nu\sb{\rm K}$=368~s if  $\nu\sb{\rm K}$/$\nu\sb{\rm v}$=3:1, and 
$\nu\sb{\rm v}$/$\nu\sb{\rm r}$=7:2. The value of P$\sb{\rm{c}}$ was fixed at its measured value. The value of 
P$\sb{\rm{v}}$ was fitted to 1085.3~s instead of the measured value of 1103.9~s, which is a 
difference of -1.7\%, and P$\sb{\rm{r}}$ was fitted to 3798.5~s instead of the measured value of 3723.2~s, which is a
difference of +2.0\%. The best-fit also provides the black hole mass and the black hole spin, which are 
 M~=~4.22$\times$10$\sp 6 M_\odot$ and  a=0.99519. The orbit resonance radius is at r=1.49, which is 1.115
times the radius
of the innermost stable circular
orbit or R~=~0.93$\times$10$\sp{12}$ cm in physical units. 
An uncertainty of $\pm$2\% in the measured values of $\nu\sb{\rm v}$ and $\nu\sb{\rm r}$ each, leads to an acceptable 
range for the black hole mass  of 4.10$\times$10$\sp 6 M_\odot$$\le$M$\le$4.34$\times$10$\sp 6 M_\odot$
and a range for the spin of 0.99472$\le${\it a}$\le$0.99561. 
 
\section{Light-curves}

According to the frequencies found temporal variations of the light-curves, both in the near-infrared as well in the X-ray band, 
should occur. 
Hot-spots are periodically generated at a rate of the vertical epicyclic frequency $\nu\sb{\rm v}$ (every 1100~s). 
The occurence rate is, however, modulated by the radial epicyclic frequency $\nu\sb{\rm r}$=2$\nu\sb{\rm v}$/7, such that 
the flare rate varies with $\nu\sb{\rm v}$$\times$(1$\pm$$\nu\sb{\rm r}$/$\nu\sb{\rm v}$). Accordingly, in snap-shot 
 like observations periods can be found which vary between 850~s and 1540~s among different observations.

Because of the resonance of $\nu\sb{\rm K}$ and $\nu\sb{\rm v}$, a hot-spot, once created, has the chance to be 
re-heated periodically, and flare again. The time separation between such flares should be exactly equal to P$\sb{\rm v}$ 
(1100~s), or 2$\times$P$\sb{\rm v}$ (2200~s).

When the flares originate from two different locations of hot-spots around the orbit, the time separation between two 
consecutive flares can differ from P$\sb{\rm v}$ by as much as one orbital period, because of the optical 
transfer function of the black hole, which causes a delay between the hot-spot creation epoch and the time when the 
image passes the line of sight of the observer. The difference between the start-times of two consecutive flares is therefore 
somewhere between 730~s$<$$\Delta$t$\sb{\rm{flare}}$$<$1470~s.

At last, the light-curves will be modulated by the orbital motion of the hot-spot, but at a frequency of 
$\nu\sb{\rm l}$ (106~s) and $\nu\sb{\rm v}$/2 (214~s). 

\section{Conclusion}

The periods indicated in earlier observations of the near-infrared and X-ray light-curves from the 
accretion disk around the Sgr A* black hole are confirmed by the analysis of the flare start-times. 
The frequencies found imply resonances of 3:1 and 7:2 suggesting that 
 radial and vertical wave oscillations as well as the Kepler frequency are involved.  
 Taking into account the rotational frame-dragging, which should 
anyhow be included in such an analysis (the Lense-Thirring effect, the Cunningham-Bardeen approach), 
the data are consistent with the presence of a supermassive,  
M~=~(4.10--4.34)$\times$10$\sp{6} M_\odot$, 
and very rapidly spinning, a~=~(0.99473--0.99561), 
black hole. The inclusion of rotational frame-dragging, which is one of the key features of 
General Relativity, is mandatory in this analysis and, of course, interpretation. 

An open issue is the ad-hoc assumption that hot-spots are created from pre-existing fluctuations. 
The analysis presented shows that the spin  
of the Sgr A* black hole is very close to ${\it a\sb{\rm{c}}}$=0.0.9953, above which the anomalous orbit
 velocity effect or 'Aschenbach' effect (Stuchl\'ik et al. 2005)
occurs (Aschenbach 2004). At ${\it a\sb{\rm{c}}}$=0.9953, the radius derivative of the 
orbital velocity $\partial{v\sp{(\Phi)}}/\partial r = 0$. This condition might cause an instability in the disk, 
like a shear-instability, which is being amplified by the radial and vertical waves.

\bibliographystyle{aa}

\vskip 1.0truecm
\noindent
{\it{Appendix}} 1

\vskip 0.5truecm
A series of start-times of flares has been measured, and the time differences $\Delta\sb{\rm i}$ 
between any two flares is searched for periodicities by scanning the values of $\Delta\sb{\rm i}$ 
with a trial period P and a best-fitting natural number, which is either  n$\sb{\rm i}$ or 
n$\sb{\rm i}$+1. The choice between n$\sb{\rm i}$ and n$\sb{\rm i}$+1 is made by the lower 
residual. There are N such intervals. 

\begin{equation}\label{eq:1}
$$a\sb{\rm i} = ( \Delta\sb{\rm i} - n\sb{\rm i}\times \rm P )\sp 2$$
\end{equation}

\begin{equation}\label{eq:2}
$$b\sb{\rm i} = ( \Delta\sb{\rm i} - (n\sb{\rm i}+1)\times \rm P )\sp 2$$
\end{equation}

\begin{equation}\label{eq:3}
$$c\sb{\rm i} = \rm{min}(a\sb{\rm i},b\sb{\rm i})$$
\end{equation}

\begin{equation}\label{eq:4}
$$X = \sqrt{{{{1}\over{N}}\sum_{i=1}^N c\sb{\rm i}}} $$
\end{equation}

The inverse of the root mean square value X divided by the 
trial period P is called 'goodness of fit' (GOF).

\begin{equation}\label{eq:5}
$$\rm{GOF} = {{\rm{P}}\over{\rm{X}}}$$
\end{equation}

\vskip 1.0truecm

\noindent
{\it{Appendix}} 2
\vskip 0.5truecm
There are three cyclic modes associated with black hole accretion disks,
which are the Kepler frequency ($\Omega\sb{\rm K}$),
the disk perturbation frequencies in vertical and radial
direction called vertical ($\Omega\sb{\rm V}$). Each  frequency
depends on the central mass M,
the angular momentum $\it a$ and the radial distance $r$ from the center.
Equations \ref{eq:1} to \ref{eq:3} show the relations 
 (Aliev \& Galtsov, 1981). 
Equation \ref{eq:4} shows the dependence on the variables for the rotational frame-dragging frequency or
$\Omega\sb{\rm{ph}}$.
The notation of c=G=1=M is used. 
Physical length scales
 are in units of GM/c$\sp 2$ and angular frequencies
$\Omega$ are in
units of c$\sp 3$/GM. $r$ = 1 is defined as the gravitational radius r$\sb{\rm g}$.

\begin{equation}\label{eq:1}
$$\Omega\sb{\rm K} = (r\sp{3/2} +  a)\sp{-1}$$
\end{equation}
\begin{equation}\label{eq:2}
$$\Omega\sb{\rm V}\sp 2 = \Omega\sb{\rm K}\sp 2 ~ (1 - {{4 a}\over{ r\sp{3/2}}} + {3 a\sp2\over{ r\sp 2}}) $$
\end{equation}
\begin{equation}\label{eq:3}
$$\Omega\sb{\rm R}\sp 2 = \Omega\sb{\rm K}\sp 2 ~ (1 - {6\over{r}} + {{8 a}\over{r\sp{3/2}}} - {3 a\sp2\over{ r\sp 2}}) $$
\end{equation}
\begin{equation}\label{eq:4}
$$\Omega\sb{\rm{ph}} = {a \over{ 2 ( 1 + \sqrt{ 1 - a\sp{2}} ) }}$$.
\end{equation}
\end{document}